# Secure Video Streaming Plug-In

Avinash Bhujbal[1], Ashish Jagtap[1], Devendra Gurav[1], Tino Jameskutty[1],

[1]Student Computer Department, ZES's DCOER, Pune

nymphjake@gmail.com

ash11.jagtap@gmail.com

dgpune@gmail.com

tinojw@yahoo.com

*Abstract* – **Video sharing sites like YouTube, Metacafe, Dailymotion, Vimeo, etc. provide a platform for media content sharing among its users. Some of these videos are copyright protected and restricted from being downloaded and saved. But users can use various download managers or application programsto download and save these videos. This affects the incoming traffic on these websites reducing their hit rate and consequently reducing their revenue. Adobe Flash Player[1]is the most commonly used player for watching online videos. It uses RTMP (Real Time Messaging Protocol)[2] to stream audio, video and data over the Internet, between a Flash Player and Adobe Flash Media Server.Here, wepropose a plug-in that enables the site owner's control over downloading of videos from such website. The plug-in will be installed at the client side with the consent of the user. When the video is being played this plug-in will send unique keys to the media server. The server will continue streaming the video after verifying the keys. Download managers or application programs will not be able to download the videos as they won't be able to create the unique keys that need to be sent to the server.**

## I. INTRODUCTION

Streaming visual data to different users is becoming increasingly popular in recent times, and protecting the transmitted data from every possible security threat has become one of the main concerns both for the end users and data providers. This paper describes a method for protecting streamed data from possible security attacks and suggests a design of secured system architecture for multimedia video streaming to multiple receivers (one receiver at a time) considering the state of the art for the video streaming existing today. The main feature of the suggested design is its ability to provide a secure communication environment for real-time data.

This downloading of videos through various streaming websites has become common scenario on the Internet.  Many sites merely warn users of copyrighted content and these users ignore the messages and use download managers or application programs to download videos. Existing video embedding methods are vulnerable to various kinds of thefts and have not been able to check piracy of contents.

The study about the different kinds of existing technologies and associated protocols used for streaming video content will provide a base to develop a plug-in that enables the site owner control over downloading of videos from his website. Ultimately the plug-in will be modified to provide security to other file formats as well.

In streaming mechanism, the file is sent to the end user in a (more or less) constant stream. It is simply a technique for transferring data such that it can be processed as a steady and continuous stream and it is called Streaming. Streaming video is a sequence of "moving images" that are sent in compressed form over the Internet and displayed by the viewer as they arrive. If a web user is receiving the videodata as streams then he/she does not have to wait to download a large file before watching the video or listening to the audio**.**

### A. Streaming Principle

Streaming multimedia is streams that is constantly received by, and normally rendered to, the end-user screen while it is being delivered by the provider. In streaming applications it is necessary for the data packets to reach their destination in a timely manner because the delay can cause the network congestion, and can result in the loss of all those packets suffering from excessive delay. This causes loss of quality of data, the synchronization between client and server to be broken, and errors to propagate in the rendered video.

There are two types of steaming, one is real time and other is prestored or prerecorded streaming. The scheme we have developed can work for both real time and prestored video files. The protocol used for streaming purpose is UDP (User Datagram Protocol)[3] which sends the media stream as a series of small packets. This is simple and efficient; however, there is no mechanism within the protocol to guarantee delivery because it does not acknowledge to the sender.

## II. LITERATURE SURVEY

There are many different implementations of existing protocols for streaming of video content over internet. Some companies have also developed their own proprietary protocol specifications.

*2.1 Existing technologies for video streaming:*

The Real Time Streaming Protocol (RTSP)[4] is a network control protocol designed for use in entertainment and communications systems to control streaming media servers. This protocol is used for establishing and controlling media sessions between end points. It has been implemented in QuickTime Streaming Server - Apple's [5] closed-source streaming server.

Real Time Messaging Protocol (RTMP) is protocol developed by Macromedia (now owned by Adobe) for streaming audio, video and data over the Internet. This protocol is used for communication between a Flash Player and Adobe Flash Media Server. RTMP sessions may be encrypted using SSL or using RTMPE, but do not provide sufficient security. Adobe Flash Player can be downloaded for free on the client side. Adobe also provides developer version of Flash Media Server for non-commercial purposes.

Also there is another technique for secure streaming which uses SSS(Secure Scalable Streaming)[6] which segments the video frames into tiles and then code tiles into header and scalable data format then packetize the header and encrypt data .

*2.2 Recent developments:*

HTML5 [7] specification provides video tag for embedding video in web pages. HTML5 video is intended by its creators to become the new standard way to show video on the web without plug-ins. Some sites like YouTube are experimenting with HTML5 video tag.

But most of the sites are not quickly adapting it due to variety of reasons – no acceptable A/V container formats; no acceptable audio and video codec, no provision for any means of encrypting the streaming or protecting the video form being downloaded absence of effective streaming protocol etc.

*2.3 Other issues involved:*

Besides all this there are also API's provided by private companies who rent their interface along with their personal players. Despite being highly priced these players still have loopholes during the buffering of videos, which can be exploited by hackers. Some media players even provide authorization after which the video can be downloaded. FlowPlayer [8] provides good interface for media content in browser and is built using Flash and JavaScript programming.

As mentioned earlier most of these players have been exploited by clever programming techniques and download managers can be used to download and save these videos. Some of these videos are copyrighted and have restricted rights of usage. Downloader applications directly connect to the server and request for video file download. These applications cause generation of unwanted traffic to the media server on sharing websites reducing their hit rate, consequently reducing the revenue.

III. PROPOSED SYSEM

The proposed solution is a browser plug-in application, which will be installed on the client with consent of the user. An application on Flash Media Server [9] will also be developed using server side Action Script.

*3.1 Requesting Video:*

When the video webpage is opened, server provides the plug-in and ensures its proper installation at client PC.

*3.2 Encrypted Unique key:*

As soon as the buffering of the video starts the client side plug-in will start sending unique keys as acknowledgements to the server. The server will check the unique keys to ensure that the video is being played and continue streaming the video.

*3.2 Algorithm:*

Analgorithm is needed for generation of keys to be sent between client and server, the Internet key exchange protocol [10] is one such method that allows two parties that have no prior knowledge of each other to jointly establish a shared secret key over an insecure communications channel. This would create a secured line of streaming of data and provide security from downloading the data.

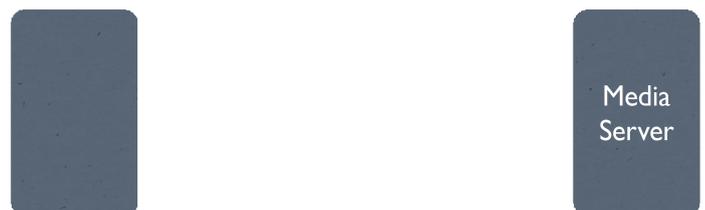

The Block diagram for the proposed system can be well explained in Fig I.

Fig I :- Block Diagram for proposed system

Thus using Flash plug-in system, the video cannot be downloaded by download managers or applicationprograms. Compatibility with other browsers can also be maintained. The plugin interface for the Flash is explained in Fig 2.

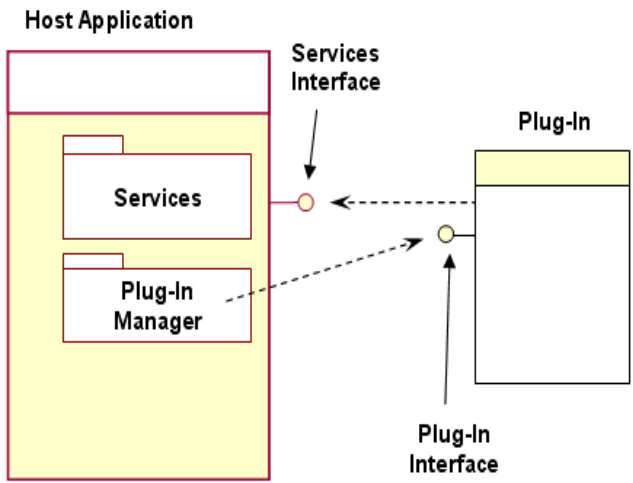

Fig 2 :- Plug-in Interface

We will use existing Adobe technologies as the base for this work as Flash has wide spread usage on the Internet. Also a plug-in developed for Flash Player will be compatible with most of the browsers. RTMP protocols can be effectively utilized for streaming video content quickly and efficiently.

Along with the plugin the Meta file obtained from the server will be encrypted using the RSA[11] which is a public key encryption system .

The RSA algorithm involves three steps: key generation, encryption and decryption.

*A. Key Generation*

RSA involves a public key and a private key**.** The public key can be known to everyone and is used for encrypting messages. Messages encrypted with the public key can only be decrypted using the private key. The keys for the RSA algorithm are generated the following way:

1. Choose two distinct prime numbers $p$ and $q$.
    - For security purposes, the integers $p$ and $q$ should be chosen at random, and should be of similar bit-length.
2. Compute $n = pq$.
    - $n$ is used as the modulus for both the public and private keys
3. Compute $\varphi(n) = (p-1)(q-1)$, where $\varphi$ is Euler's totient function.
4. Choose an integer $e$ such that $1 < e < \varphi(n)$ and greatest common divisor of $(e, \varphi(n)) = 1$; i.e., $e$ and $\varphi(n)$ are coprime.
    - $e$ is released as the public key exponent.
    - $e$ having a short bit-length and small Hamming weight results in more efficient encryption . However, small values of $e$ (such as 3) have been shown to be less secure in some settings.
5. Determine $d$ as:

$$d \equiv e^{-1} \pmod{\varphi(n)}$$

i.e., $d$ is the multiplicative inverse of $e$ mod $\varphi(n)$.

- This is more clearly stated as solve for d given (de) = 1 mod $\varphi(n)$
- This is often computed using the extended Euclidean algorithm.
- $d$ is kept as the private key exponent.

By construction, d*e= 1 mod $\varphi(n)$. The public key consists of the modulus $n$ and the public (or encryption) exponent $e$. The private key consists of the modulus $n$ and the private (or decryption) exponent $d$ which must be kept secret. ($p$, $q$, and $\varphi(n)$ must also be kept secret because they can be used to calculate $d$.)

*B. Encryption*

'A' transmits her public key $(n, e)$ to 'B' and keeps the private key secret. Bob then wishes to send message **M** to Alice.

He first turns **M** into an integer m, such that $0 \leq m < n$ by using an agreed-upon reversible protocol known as a padding scheme. He then computes the cipher text $c$ corresponding to

$$c = m^e \pmod{n}$$

This can be done quickly using the method of exponentiation by squaring. 'B' then transmits $c$ to 'A'.

*C. Decryption*

'A' can recover $m$ from $c$ by using her private key exponent $d$ via computing

$$m = c^d \pmod{n}$$

Given $m$, she can recover the original message **M** by reversing the padding scheme.

IV. CONCLUSION

In this paper, we have presented one method for protecting streaming video content. This method can be implemented in Flash Player or in other products available for media content delivery over Internet. It is important to understand that more secure the key generation algorithm is, the more restrictive the ability to break this system and download video content will become. But the video content is still vulnerable to copy attacks at low level graphics processing unit at client side.

This approach adds a measure of security but is primarily intended to inhibit or reduce the threat of video content piracy. As with any security model, a dedicated hacker over time can successfully find a way around these measures. The method presented work will successfully obstruct most potential threats or reduce it significantly.